\documentclass[aps,prb,floatfix,epsfig,twocolumn,showpacs,preprintnumbers]{revtex4}
\usepackage{amsfonts}
\usepackage{amssymb}
\usepackage{graphicx}
\usepackage{amsmath}
\usepackage{tablefootnote}
\usepackage{setspace} 

\begin{document}

\title{Self-consistent solution of Hedin's equations: semiconductors/insulators}
\author{Andrey L. Kutepov\footnote{e-mail: kutepov@physics.rutgers.edu}}
\affiliation{Department of Physics and Astronomy, Rutgers University, Piscataway, NJ 08856}

\begin{abstract}
The band gaps of a few selected semiconductors/insulators are obtained from the self-consistent solution of the Hedin's equations. Two different schemes to include the vertex corrections are studied: (i) the vertex function of the first-order (in the screened interaction $W$) is applied in both the polarizability $P$ and the self-energy $\Sigma$, and (ii) the vertex function obtained from the Bethe-Salpeter equation is used in $P$ whereas the vertex of the first-order is used in $\Sigma$. Both schemes show considerable improvement in the accuracy of the calculated band gaps as compared to the self-consistent $GW$ approach (sc$GW$) and to the self-consistent quasi-particle $GW$ approach (QS$GW$). To further distinguish between the performances of two vertex-corrected schemes one has to properly take into account the effect of the electron-phonon interaction on the calculated band gaps which appears to be of the same magnitude as the difference between schemes i) and ii).
\end{abstract}

\pacs{71.20.Mq, 71.20.Nr, 71.20.Ps, 71.45.Gm}
\maketitle

\section*{Introduction}
\label{intro}

\begin{figure}[t]
\includegraphics[width=7.0 cm]{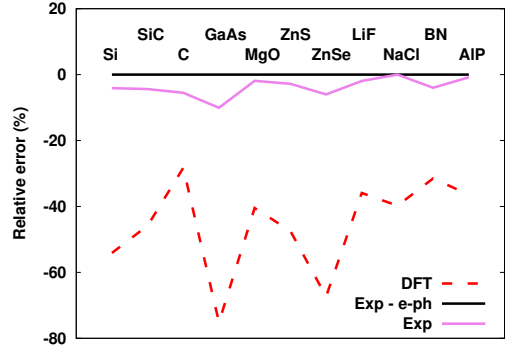}
\caption{(Color online) Relative errors of the calculated (in GGA) band gaps of selected semiconductors/insulators. Experimental data have been taken from the Refs.[\onlinecite{prb_20_624,prb_35_9174,prb_53_16283}]. Horizontal line represents the experimental data corrected by the effects of the electron-phonon and the spin-orbit interactions. The corrections were taken from the Refs.[\onlinecite{prb_89_214304,prl_112_215501,rmp_77_1173,prb_93_100301}].}
\label{dft_gaps}
\end{figure}

The ability to accurately predict the band gaps in semiconductors/insulators is a long-standing goal for computational physicists. Density functional theory\cite{pr_140_1133} (DFT) in its Local Density Approximation (LDA) or in Generalized Gradient Approximation (GGA) has been very successful in the prediction of the ground state properties, but its application to the electronic structure problem has been a failure. Specifically, the average error in the calculated band gaps of semiconductors/insulators is about 40-50\% as can be estimated from Fig.\ref{dft_gaps}. The error becomes even larger if one takes into account the effect of the electron-phonon (e-ph) interaction. Common approach nowadays to address the band gap problem, therefore, is to use Hedin's $GW$ approximation\cite{pr_139_A796} or its numerous extensions\cite{prb_25_2867,prb_34_5390,prb_37_10159,prl_59_819,prl_99_246403,prl_89_126401,prb_76_165106,prb_85_155129,
prb_57_11962,prl_88_066404,prl_112_096401} which often are more accurate than DFT in this respect. A great variety of different $GW$-based schemes, which are available today, allows one to select the most accurate one for a given material. On the other hand, the same variety of approaches tells us that the search for an optimal method to calculate the band gaps has not been finished yet. As it was discussed in the recent paper (Ref.[\onlinecite{prb_94_155101}], which hereafter will be abbreviated as I), the success of many existing extensions of the $GW$ approach is based on the cancellation of errors which renders their systematic improvement complicated. Besides, $GW$ schemes do not provide an information on the effect of the diagrams beyond $GW$. For example, one cannot say whether the perturbation expansion is convergent or not. Common route to go beyond GW approximation nowadays is to combine the GW approach with the Time Dependent Density Functional Theory (TDDFT) through the introduction of the exchange-correlation kernel $f_{xc}$.\cite{prb_49_8024,prl_88_066404,prl_94_186402} But to the best of my knowledge, until recently, there were no attempts to apply self-consistent purely many-body extensions of the GW method to the crystalline materials.

In I, a few schemes to solve the Hedin's equations\cite{pr_139_A796,rdnc_11_1} self-consistently (sc) and with higher order diagrams (vertex corrections) included have been introduced. The schemes are not based on the cancellation of errors. The main advantages of them as compared to the existing $GW$-based methods are the following: they are diagrammatic and self-consistent, they do not apply the
quasi-particle approximation for the Green's function, they treat full frequency dependence of the interaction $W$ in higher order diagrams, they apply the vertex corrections for both the polarizability and the self-energy. The schemes were successfully used in I to calculate the band widths in alkali metals (Na and K) and the band gaps in Si and LiF. The accuracy of the obtained results was superior to the accuracy of the QS$GW$ method and it was generally within uncertainty of the experimental data, but the calculations presented in I were very time consuming.

In this work, a few improvements of the algorithms presented in I are used, which makes the approaches less computationally demanding. The improvements continue to explore the fact that the diagrams beyond $GW$ (vertex part) can generally be accurately evaluated with less intensive numerical parameters as compared to the parameters one uses in the $GW$ part. This idea was partially used already in I. Namely, the number of band states ($N_{bnd}$) (they are obtained from the effective Hartree-Fock problem\cite{prb_85_155129} on every iteration) used in the vertex part was less than in the $GW$ part. Also, the number of orbitals ($\phi_{nl}$) inside the MT (muffin-tin) spheres and the number of plane waves in the interstitial region ($N_{\mathbf{G}}$) to represent the band states were smaller in the vertex part. In this work, the coarser \textit{k}-mesh in the Brillouin zone (with $N_{\mathbf{k}}$ points) and a smaller number of the Matsubara's time/frequency points ($n_{\tau}$/$n_{\omega}$) for the vertex part have been used.

The plan of the paper is the following. Section \ref{meth} explains the selection of the vertex corrected schemes for this study and presents the convergence checks. Section \ref{res} provides the
results obtained and the discussion. The conclusions are given afterwords.

\section{Methods and convergence checks}
\label{meth}

\begin{figure}[b]
\includegraphics[width=5.0 cm]{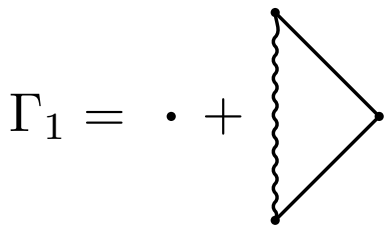}
\caption{First order approximation for the 3-point vertex function. Direct lines represent Green's function G and the wavy line represents the screened interaction W. The dot is for the trivial part of the vertex function.}
\label{gamma_1}
\end{figure}

In this work, for a systematic study of the band gap problem in semiconductors, two of the six approaches introduced in I are used: B and D schemes. Scheme B is conserving and represents the first step beyond GW approximation where all diagrams up to the first order (in the screened interaction W) are included in the vertex function (Fig.\ref{gamma_1}). The same vertex function ($\Gamma=\Gamma_{1}$) is then applied for both the polarizability

\begin{equation}\label{def_pol1}
P(12)=\sum_{\alpha}G^{\alpha}(13)\Gamma^{\alpha}(342)G^{\alpha}(41),
\end{equation}

and the self-energy
 
\begin{equation}\label{def_M8}
\Sigma^{\alpha}(12)= - G^{\alpha}(14)\Gamma^{\alpha}(425)W(51),
\end{equation}
where digits inside the brackets represent the space-time arguments and $\alpha$ is the spin index. All relevant quantities (P, W, G, $\Sigma$) are iterated till the full self-consistency in scheme B. The satisfaction of the macroscopic conservation laws is appealing, but the limited number of the diagrams included in scheme B breaks down other relationships which should be fulfilled in an exact theory. For instance, the polarizability in scheme B doesn't match its natural definition as a functional derivative of the electron density with respect to the electric field (external plus induced). Thus, the polarizability in scheme B is not physical.

\begin{figure}[t]
\includegraphics[width=8.0 cm]{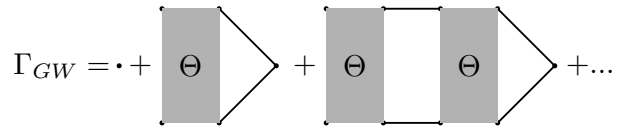}
\caption{Ladder sequence for the 3-point vertex function with the kernel
$\Theta$ (see Fig.\ref{theta_appr}) as the rung of the ladder.} \label{gamma_theta}
\end{figure}

\begin{figure}[b]
\includegraphics[width=7.0 cm]{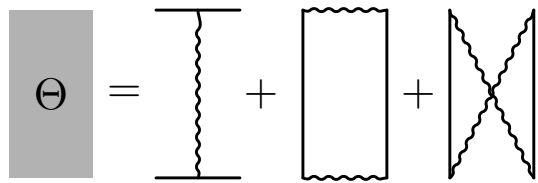}
\caption{The GW approximation for the irreducible 4-point kernel $\Theta$. Direct lines represent the Green's function and wavy lines represent the screened interaction W.} \label{theta_appr}
\end{figure}

\begin{table}[t]
\caption{Dependence of the calculated band gap of MgO on the calculation setup for the diagrams beyond $GW$. Scheme B (see the text for the details) has been used. $\phi_{nl}$ enumerates the orbitals (shown as sequences of principal quantum numbers and orbital characters) inside MT spheres which were used to represent the band states in the diagrams beyond GW approximation.} \label{conv}
\begin{center}
\begin{tabular}{@{}c c c}  Parameter & Setup & Band gap\\
\hline\hline
$N_{bnd}$ & 0 &9.31 \\
 & 5 &8.81\\
 & 10 &8.40 \\
 & 20 &8.28 \\
 & 30 &8.29 \\
\hline
$\phi_{nl}$ &3s(Mg)/2p(O) &8.56 \\
 &3s3p(Mg)/3s2p(O) &8.40 \\
 &3s3p3d(Mg)/3s2p3d(O) &8.28 \\
 &3s3p3d4f(Mg)/3s2p3d4f(O) &8.27 \\
\hline
$N_{\mathbf{G}}$ & 26 &8.19\\
 & 59 &8.25\\
 & 92 &8.28\\
\hline
$N_{\mathbf{k}}$ & $2^{3}$ &8.28 \\
 & $3^{3}$ &8.24 \\
 & $4^{3}$ & 8.27\\
\hline
$n_{\tau}$, $n_{\omega}$, $n_{\nu}$
  & 46 &8.28\\
  & 62 &8.29\\
  & 94 &8.29
\end{tabular}
\end{center}
\end{table}

\begin{figure}[t]
\centering
\includegraphics[width=9.0 cm]{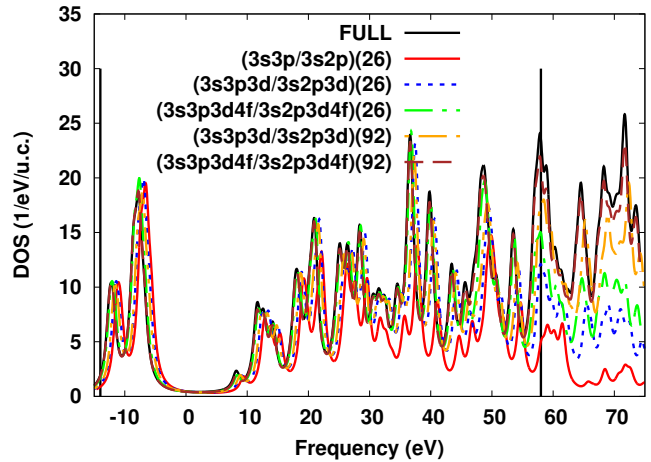}
\caption{(Color online) Dependence of the density of states of MgO (from the effective Hartree-Fock problem) on the quality of the representation of the band states. Line 'FULL' corresponds to the properly normalized band states (in this case they are represented with the sets of plane waves and $\phi$'s as in $GW$ part, see Table \ref{setup}). The rest of the lines have been obtained with the sets of plane waves and $\phi$'s of different completeness. Each line is marked with the corresponding set of $\phi$'s (first entry, Mg/O), and the number of plane waves $N_{\mathbf{G}}$ in the interstitial region (second entry). Two vertical lines show the range of the band states ($\sim$20 bands) used in the vertex part as a basis set. Formula (D4) from Ref.[\onlinecite{prb_85_155129}] has been used to evaluate the spectral function. Smearing parameter was set to 0.005 Ry to avoid too sharp peaks. The \textit{k}-mesh $3\times3\times3$ corresponding to the vertex part was used.} \label{sf_mt}
\end{figure}

\begin{figure}[b]
\centering
\includegraphics[width=9.0 cm]{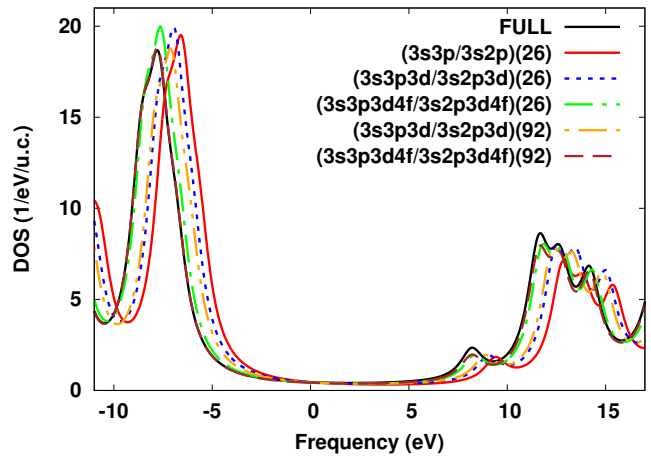}
\caption{(Color online) The same as Figure \ref{sf_mt} but for low energy part of the spectrum.} \label{sf_mt_low}
\end{figure}

On the other hand, the polarizability in scheme D is physical by construction, as the corresponding vertex function ($\Gamma=\Gamma_{GW}$) is obtained from the Bethe-Salpeter equation (Fig.\ref{gamma_theta}) with the kernel ($\Theta=\frac{\delta \Sigma}{\delta G}$) in "GW" approximation (Fig.\ref{theta_appr}). Thus, in scheme D, the polarizability (and the corresponding W) are not iterated till the full self-consistency but are evaluated only once, immediately after scGW calculation. The choice of the diagrams for the self energy in scheme D (equation (\ref{def_M8}) with $\Gamma=\Gamma_{1}$) is a trade between the consistency with the Bethe-Salpeter equation for the polarizability on one hand (the kernel of the Bethe-Salpeter equation is obtained assuming $\Sigma=GW$) and the explicit "improvement" of the self-energy itself by including higher order skeleton diagrams on the other hand. The self energy and Green's function are iterated till full self-consistency in scheme D, similar to scheme B. Scheme D represents an example of a diagrammatic approach where certain infinite series of diagrams are included based on a specific physical principle (in this particular case - the requirement of the microscopic charge preservation).

It is important to mention that the frequency dependence of W is taken into account without approximations in both schemes B and D. This allows one to consider them as advanced approaches as compared to the commonly used static approximation for W (taken at zero frequency) or the plasmon-pole approximation.\cite{prb_34_5390} One more technical detail is of importance: in the diagrams beyond "GG" approximation for the polarizability and "GW" approximation for the self-energy all building blocks (G and W) are treated with the same (reduced) basis set. Thus, no asymmetry is introduced in the self-energy from the terms like $G*W*\Gamma[G,W]$ because all G and W (not only those which are the arguments of the $\Gamma$) are expressed in the same reduced basis set.

As the above two vertex-corrected schemes represent different approaches for the selection of diagrams, it seems to be interesting to apply them systematically for semiconducting/insulating materials. For the comparative purposes, the calculations with scGW approach (scheme A) and with QSGW approach have also been performed in this study.

An important step is the checking that the band gaps are converged with respect to the calculation setup for the vertex part, which was less intensive as compared to the setup for the GW part. Table \ref{conv} shows how the evaluated band gap of MgO (all studied in this work materials show similar convergence) depends on the calculation setup for the vertex part. The setup for the $GW$ part was fixed at the level presented below in the Table \ref{setup}. When a certain setup parameter (in the vertex part) was varied, the rest of the vertex part parameters were kept at their values also indicated in Table \ref{setup}. As it follows, the most sensitive parameter is the number of the band states included in the basis set. However, with about 20-30 bands being sufficient for the vertex part, this number is approximately seven times less than the number of the band states needed for the GW part (Table \ref{setup}). One can estimate from Table \ref{conv} that the uncertainty of the calculated band gap related to the changing of the vertex part setup is about $0.03 eV$, which comprises less than 5\% of the vertex correction to the band gap (see Table \ref{b_gaps} below). It is also a few times smaller than the difference between the band gaps obtained in schemes B and D. Thus, the accuracy is sufficiently good to allow us to compare the different schemes (at least at the level when one neglects by the e-ph interaction).

\begin{table*}[t]
\caption{Setups of the calculations. $N_{\mathbf{k}}$ stands for the number of \textbf{k}-points in the Brillouin zone. $N_{lapw}$ is the number of linearized augmented plane waves (LAPW's) in the basis set ($GW$ part). $\phi_{nl}$ enumerates all orbitals (shown as sequences of principal quantum numbers and orbital characters) inside MT spheres which were used to represent the band states. Those of them shown in bold were used to augment the plane waves. The rest are the local orbitals (LO's). The total number of bands used as a basis for $GW$ part of the calculations is the sum of all LO's and LAPW's. The columns $(L_{max}/N)_{MT}^{PB}$ show maximal orbital character ($L_{max}$) and total number of Product Basis (PB) functions in the MT sphere for a specific atom. $N_{\mathbf{G}}^{PB}$ stands for the number of PB functions in the interstitial region ($GW$ part). $N_{\mathbf{G}}$ shows (for the Vertex part) the number of plane waves used to represent the band states and the PB functions in the interstitial region. The number of points on the Matsubara's time/frequency meshes was 62 in the $GW$ part and 46 in the vertex part. The basis set of the band states in the vertex part included 20-30 bands closest to the chemical potential.} \label{setup}
\begin{center}
\singlespacing
\resizebox{\textwidth}{!}{
\begin{tabular}{@{}c c c c c c c c c c}  &\multicolumn{5}{c}{GW part} &
\multicolumn{4}{c}{Vertex part}\\
 & $N_{\mathbf{k}}$ & $N_{lapw}$  & $\phi_{nl}$  & $(L_{max}/N)_{MT}^{PB}$ & $N_{\mathbf{G}}^{PB}$   & $N_{\mathbf{k}}$& $\phi_{nl}$& $(L_{max}/N)_{MT}^{PB}$ & $N_{\mathbf{G}}$\\
\hline
Si & $8^{3}$ & 190 & \textbf{3s3p3d4f5g6h7i}2s4s5s2p4p5p4d5d5f6g7h(Si) &6/332(Si)   & 652 &$4^{3}$  &\textbf{3s3p}(Si)  & 2/19(Si) &28 \\
 &  &  &\textbf{1s2p3d4f5g6h7i}2s3p4d(Em)  &6/236(Em)   &  &  &\textbf{1s2p}(Em)  &2/27(Em) & \\
SiC &$6^{3}$  &144  &\textbf{3s3p3d4f5g6h7i}2s4s2p4p4d5f6g(Si)  &6/294(Si)   &650  &$3^{3}$  &\textbf{3s3p}(Si)  &2/19(Si) &59 \\
 &  &  &\textbf{2s2p3d4f5g6h7i}3s4s3p4p4d5f6g(C)  &6/291(C)   &  &  &\textbf{2s2p} (C) &2/19(C) & \\
 &  &  &\textbf{1s2p3d4f5g6h}2s3s3p4d5f(Em)  &6/261(Em)   &  &  &\textbf{1s2p}(Em)  &2/27(Em) & \\
C &$8^{3}$  &144  &\textbf{2s2p3d4f5g6h7i}1s3s4s3p4p4d5f(C) &6/275(C)   &652  &$4^{3}$  &\textbf{2s2p}(C)  &2/27(C) &28 \\
 &  &  &\textbf{1s2p3d4f5g6h7i}2s3p4d(Em)  &6/236(Em)   &  &  &\textbf{1s2p}(Em)  &2/27(Em) & \\
GaAs &$6^{3}$  &144  &\textbf{4s4p4d4f5g6h7i}3s5s3p5p3d5d5f6g(Ga)  &6/344(Ga)   &650  &$3^{3}$  &\textbf{4s4p4d}(Ga)  &4/89(Ga) &59 \\
 &  &  &\textbf{4s4p4d4f5g6h7i}3s5s3p5p3d5d5f6g(As)  &6/344(As)   &  &  &\textbf{4s4p4d}(As) &4/72(As) & \\
 &  &  &\textbf{1s2p3d4f5g6h7i}2s3s3p4d5f(Em)  &6/261(Em)   &  &  &\textbf{1s2p}(Em)  &2/27(Em) & \\
MgO &$6^{3}$  &113  &\textbf{3s3p3d4f5g6h7i}2s4s5s2p4p5p4d5d5f6g7h(Mg) &6/332(Mg)   &380  & $3^{3}$ &\textbf{3s3p3d}(Mg)  &4/97(Mg) &59 \\
 &  &  &\textbf{3s2p3d4f5g6h7i}2s4s3p4p4d5d5f6g7h(O)  &6/314(O)   &  &  &\textbf{3s2p3d}(O) &4/90(O) & \\
ZnS &$6^{3}$  &190  &\textbf{4s4p3d4f5g6h7i}3s5s6s3p5p6p4d5d6d5f6f6g7h(Zn)  &6/381(Zn)   & 650 &$3^{3}$  &\textbf{4s4p}(Zn)  &2/19(Zn) &59 \\
 &  &  &\textbf{3s3p3d4f5g6h7i}2s4s5s2p4p5p4d5d5f6g7h(S)  &6/292(S)   &  &  &\textbf{3s3p}(S) &2/19(S) & \\
 &  &  &\textbf{1s2p3d4f5g6h7i}2s3s3p4p4d5f6g7h(Em)  &6/271(Em)   &  &  &\textbf{1s2p}2s(Em)  &2/27(Em) & \\
ZnSe &$6^{3}$  &190  &\textbf{4s4p3d4f5g6h7i}3s5s6s3p5p6p4d5d6d5f6f6g7h(Zn)  &6/381(Zn)   &648  &$3^{3}$  &\textbf{4s4p}(Zn)  &2/19(Zn) & 59\\
 &  &  &\textbf{4s4p4d4f5g6h7i}3s5s6s3p5p6p3d5d6d5f6g7h(Se)  &6/226(Se)   &  &  &\textbf{4s4p} (Se) &2/19(Se) & \\
 &  &  &\textbf{1s2p3d4f5g6h7i}2s3s3p4p4d5f6g7h(Em)  &6/271(Em)   &  &  &\textbf{1s2p}(Em)  &2/27(Em) & \\
LiF &$6^{3}$  &113  &\textbf{2s2p3d4f5g6h7i}1s3s4s3p4p4d5f6g7h(Li)  &6/286(Li)   &376  &$3^{3}$  &\textbf{2s2p}(Li)  &2/28(Li) &59 \\
 &  &  &\textbf{2s2p3d4f5g6h7i}3s4s3p4p4d5f6g7h(F)  &6/286(F)   &  &  &\textbf{2s2p3d}(F) &4/81(F) & \\
NaCl &$6^{3}$  &113  &\textbf{3s3p3d4f5g6h7i}2s4s5s2p4p5p4d5d5f6g7h(Na)  &6/332(Na)   &376  &$3^{3}$  &\textbf{3s3p}(Na)  &2/19(Na) &59 \\
 &  &  &\textbf{3s3p3d4f5g6h7i}4s5s4p5p4d5d5f6g7h(Cl)  &6/329(Cl)   &  &  &\textbf{3s3p}(Cl) &2/19(Cl) & \\
BN &$6^{3}$  &113  &\textbf{2s2p3d4f5g6h7i}1s3s3p4d(B) &6/255(B)   &380  &$3^{3}$  &\textbf{2s2p}(B)  &2/28(B) &59 \\
 &  &  &\textbf{2s2p3d4f5g6h7i}3s3p4d(N)  &6/255(N)   &  &  &\textbf{2s2p}(N) &2/19(N) & \\
 &  &  &\textbf{1s2p3d4f5g6h7i}2s3p(Em)  &6/222(Em)   &  &  &\textbf{1s2p}(Em)  &2/27(Em) & \\
AlP &$6^{3}$  & 190 &\textbf{3s3p3d4f5g6h7i}2s4s2p4p4d5f(Al)  &6/290(Al)   &650  &$3^{3}$  &\textbf{3s3p}(Al)  & 2/19(Al) &59 \\
 &  &  &\textbf{3s3p3d4f5g6h7i}2s4s2p4p4d5f(P)  &6/290(P)   &  &  &\textbf{3s3p}(P) &2/19(P) & \\
 &  &  &\textbf{1s2p3d4f5g6h7i}2s3p(Em)  &6/222(Em)   &  &  &\textbf{1s2p}(Em)  &2/27(Em) & 
\end{tabular}}
\end{center}
\end{table*}

As it was mentioned above, the basis set for the vertex part consists from the smaller number of band states and the representation of these "lower energy" band states inside the MT spheres and in the interstitial region is provided with a reduced number of the $\phi_{nl}$-orbitals and the plane waves correspondingly. So, it is interesting to check how this approximate representation for the "lower energy" band states affects their normalization and how it is related to the convergence of the calculated band gap presented in the Table \ref{conv}. A convenient way to compare the normalization of the band states is to plot the density of states (DOS) (Figures \ref{sf_mt} and \ref{sf_mt_low}). As one can see, the accuracy of the representation is very good at the energies close to the chemical potential. The accuracy is under control in the interval of the band states used as a basis in the vertex part of the calculations. The most important $\phi_{nl}$-orbitals for the vertex part are the $3s$-orbitals of Mg and $2p$-orbitals of O. They ensure proper normalization of the band states at low energies and they provide most of the band gap correction (see Table \ref{conv}). $spd$-basis set in the MT spheres and 59 plane waves in the interstitial region allow to obtain well converged band gaps for this material. Inclusion of the f-orbitals inside the MT spheres and further increasing the number of plane waves in the interstitial region improve the normalization at higher energies (Fig.\ref{sf_mt}), but do not affect the calculated band gap much, providing a further support that the selected basis set (Table \ref{setup} below) for the vertex part is sufficiently good for our purposes.

\begin{table*}[t]
\caption{Theoretical and experimental band gaps (eV). Abbreviations A, B, and D have been introduced in I. Experimental data have been cited from Refs.[\onlinecite{prb_20_624,prb_76_165106,prb_35_9174,prb_53_16283}]. Corrected experimental results (electron-phonon/spin-orbit interaction) are based on Refs.[\onlinecite{prb_89_214304,prl_112_215501,rmp_77_1173,prb_93_100301}].} \label{b_gaps}
\begin{center}
\begin{tabular}{@{}c c c c c c c c c c}  &\multicolumn{3}{c}{Earlier QS$GW$} &
\multicolumn{4}{c}{Present work} & exp  & exp+corr\\
 & Ref.[\onlinecite{prb_76_165106}] &Ref.[\onlinecite{prl_99_246403}]  &Ref.[\onlinecite{prb_92_041115}]   &  QS$GW$ & A & B& D &  & \\
\hline
Si & 1.23 & 1.41 & 1.47 & 1.41 & 1.55 &1.32  &1.26  &1.17  &1.22-1.24\\
SiC &2.52  &2.88  &2.90  &2.79  &2.89  &2.52  &2.42  &2.40  &2.51 \\
C &5.94  &6.18  &6.40   &6.18  &6.15  &5.80  &5.73  &5.48  &5.80-5.88 \\
GaAs &1.93  &1.85  &1.75   &1.96  &2.27  &1.80  &1.72  &1.52  &1.69 \\
MgO &  &9.16  & 9.29  &9.42  &9.31  &8.24  &7.96  &7.83  &7.98 \\
ZnS &4.04  &4.15  &  &4.19  &4.28  &3.90  &3.79  &3.83  &3.94 \\
ZnSe &3.08  &  &     &3.17  &3.32  &2.96  &2.80  &2.82  &3.00 \\
LiF &  &15.9  &     &16.63  &16.30  &15.02  &14.39  &14.2  &14.48 \\
NaCl &  &  &    &9.81  &9.25  &8.55  &8.49  &8.5  & \\
BN &  &7.14  &7.51  &7.06  &7.06  & 6.37 &6.30  &6.25  &6.51,6.6 \\
AlP &  & 2.90 &3.10    &2.80  &2.84  &2.53  &2.44  &2.45  & 2.47
\end{tabular}
\end{center}
\end{table*}

For the materials with a band gap, studied in this work, the quality of the \textbf{k}-mesh in the Brillouin zone is not very critical and, correspondingly, it was safe to use the coarser meshes like $3\times 3\times 3$ or even $2\times 2\times 2$ for the vertex part. In the case of metals, the using of the coarser meshes for the vertex part should still be appropriate with the exception of when the vertex function is evaluated for zero (or very small) external bosonic frequency. In this case, one has to make sure that the details of the Fermi surface are handled with sufficient accuracy.

Table \ref{setup} shows the setups of the calculations for all materials studied in this work, distinguishing the basis sets used in the $GW$ part and in the vertex part. The information presented in Table \ref{setup} might be useful as a reference in the future. Experimental crystal structures have been used: the diamond structure for Si and C, the zinc-blende structure for SiC, GaAs, ZnS, ZnSe, BN, AlP, and the rocksalt (B1) structure for LiF, NaCl. Empty spheres (Em) were used for the materials with diamond and zinc-blende crystal structures to enhance the accuracy of the interstitial region description. Temperature was fixed at 1000K.

\section{Results}
\label{res}

The main results of this work are collected in Table \ref{b_gaps}, where the band gaps obtained with two vertex corrected schemes (B and D in the notations of I) are compared with sc$GW$ (scheme A) results as well as with the earlier and present QS$GW$ calculations and with the experimental data. The principal conclusion from Table \ref{b_gaps} is the following: vertex corrected schemes (B and D) allow one to considerably improve the results from sc$GW$/QS$GW$ calculations. As one can see, the biggest correction comes from the first order vertex (scheme B), and scheme D reduces the gaps from scheme B a little more. Essentially, the difference between the band gaps obtained with schemes B and D (and the deviation of the corresponding gaps from the experimental data) is of about the same magnitude or less than the correction of the band gaps originating from the electron-phonon/spin-orbit interaction (MgO and LiF slightly deviate from this rule). The fact that the effects of the electron-phonon interaction were evaluated only approximately (often with use of the LDA wafe functions to evaluate the correction) makes it difficult to conclude decisively wchich scheme (B or D) is better. Thus, further improvement in the theoretical approach should, obviously, include electron-phonon interaction in a certain way.

QS$GW$ calculations in this work have shown only marginal (if any) improvement as compared to the sc$GW$ approach and the corresponding band gaps differ noticeably from the experimental data. Among the QS$GW$ results published earlier, the band gaps reported in Ref.[\onlinecite{prb_92_041115}] are the closest ones to the QS$GW$ gaps obtained in this study. There are considerable differences in the gaps reported in the earlier QS$GW$ works. Particularly, the gaps from Ref.[\onlinecite{prb_76_165106}] are systematically smaller (GaAs is an exception) than the gaps reported in Refs.[\onlinecite{prl_99_246403},\onlinecite{prb_92_041115}] and obtained in this study. Differences in numerical methods (linear muffin-tin orbitals (LMTO) in [\onlinecite{prb_76_165106}], projector-augmented-wave (PAW) in [\onlinecite{prl_99_246403}], and norm-conserving pseudo-potentials in [\onlinecite{prb_92_041115}]) can contribute to the discrepancies among calculated band gaps. Besides, the construction of the quasiparticle spectrum in [\onlinecite{prl_99_246403}] is somewhat different. The disagreements may also arise from the different degree of convergence with respect to the basis set. This speculation appears when one thinks about a rather strong effect of the high energy local orbitals (LO) in the LAPW+LO basis set upon the calculated band gaps reported recently in $G_{0}W_{0}$ calculations.\cite{prb_74_045104,prb_83_081101,prb_84_039906,prb_93_115203,prb_94_035118}

\begin{figure}[b]
\centering
\includegraphics[width=9.0 cm]{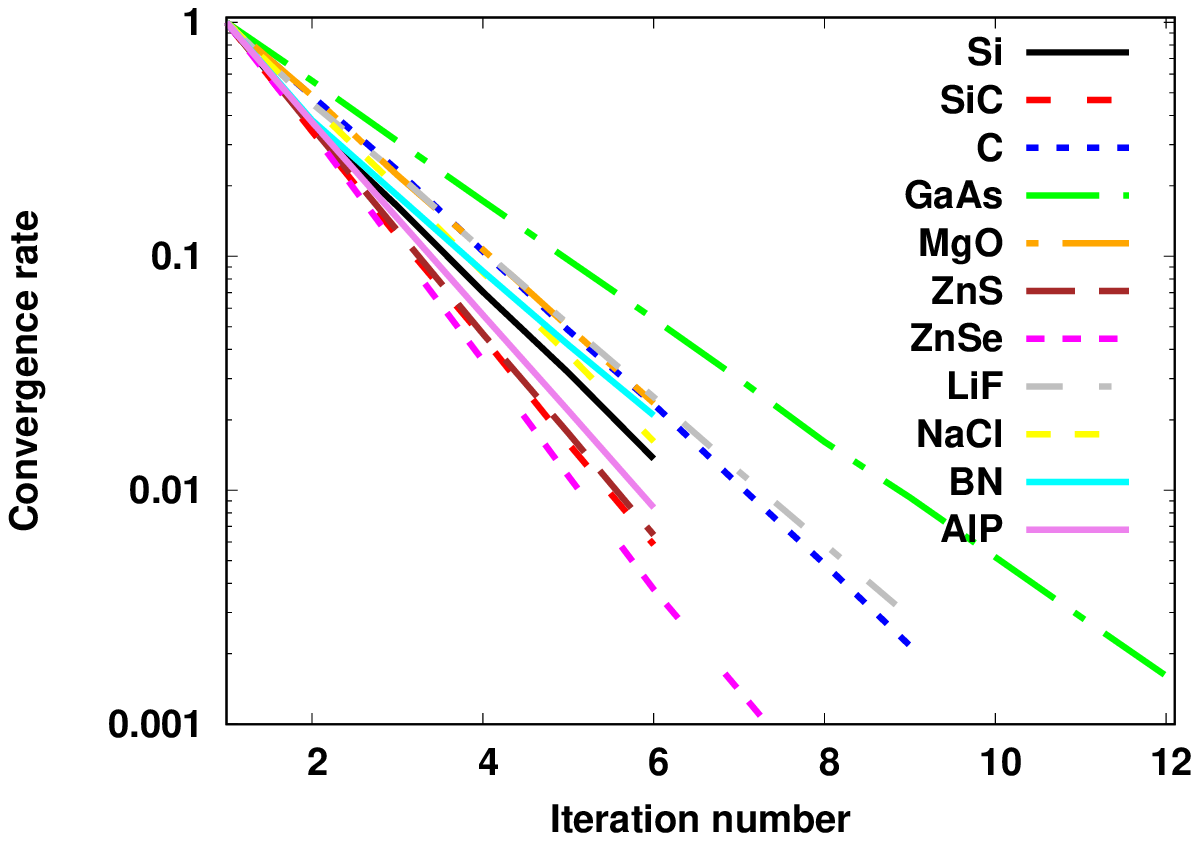}
\caption{(Color online) Convergence of the vertex function with respect to the iteration number when one solves the Bethe-Salpeter equation. Shown are the ratios of the largest element of the correction to the vertex at a given iteration and at the first iteration.} \label{bs_conv}
\end{figure}

Analyzing the calculations performed with scheme D, one can learn about the convergence of the diagrammatic series. Namely, every subsequent iteration of the Bethe-Salpeter equation represents an addition of the more complicated diagram to the vertex function. Figure \ref{bs_conv} shows the convergence of this process. All studied materials show very good convergence. After six iterations (9 in case of GaAs), the correction is reduced by two orders, which is enough for a very good convergence of the band gaps. The reason for a somewhat slower convergence in the case of GaAs is not clear at this point and can be an objective of a separate study.

\begin{figure}[t]
\centering
\includegraphics[width=9.0 cm]{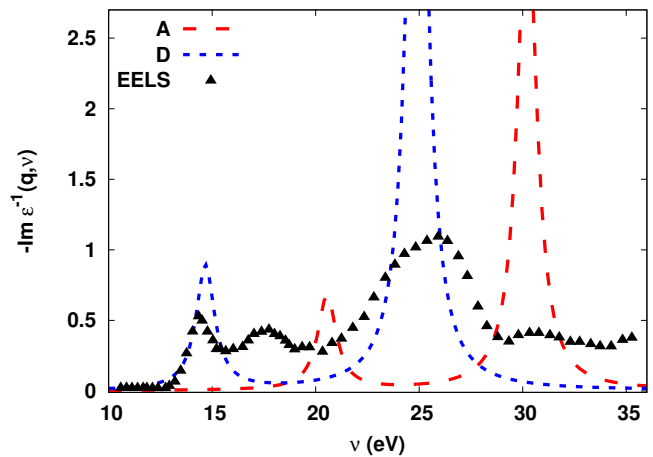}
\caption{(Color online) Calculated electron energy loss spectrum of LiF at q=0.5GX. Symbols represent the EELS data reproduced from the Ref.[\onlinecite{prl_84_3907}]} \label{eels}
\end{figure}

Whereas the principal goal of this study was to clarify the effect of vertex corrections on the calculated band gaps, it is obvious that other information can be extracted from the calculations. As an example, Fig.\ref{eels} shows the imaginary part of the inverse dielectric function $\epsilon^{-1}(\mathbf{q},\nu)$ as a function of real frequency $\nu$ evaluated for the momentum $\mathbf{q}=0.5\Gamma X$ (halfway between $\Gamma$ and $X$ points in the Brillouin zone). The calculated results are compared with the experimentally measured electron energy loss spectrum (EELS). Presented theoretical curves were obtained by analytically continuing the $\epsilon^{-1}(\mathbf{q},i\nu)$ 
calculated as a function of Matsubara's frequency. As one can see, by taking into account the multiple electron-hole scattering through the solution of the Bethe-Salpeter equation (scheme D) one can considerably improve the agreement with the experimental data. Two principal peaks at approximately 15eV and 25eV are reproduced sufficiently well. The smaller peak at about 18eV noticeable in the experimental data is absent in the calculations, presumably because of the insufficient accuracy of the analytical continuation. Another drawback of the need to perform the analytical continuation is that the original data (on the imaginary axis) have to be of a very high quality. Particularly, the function $\epsilon^{-1}(\mathbf{q},\nu)$ with $\nu$ on the real axis was stabilized only when the number of Matsubara's time points was increased up to 100-120. It is interesting, that the band gaps, the evaluation of which also includes analytical continuation (of the self-energy), show considerably faster convergence (Table \ref{conv}).

Of course, the calculation of the dielectric function and related with it properties by itself does not represent anything new in the computational solid state physics. In fact, Bethe-Salpeter equation as a tool for studying the excitons was introduced more than 50 years ago\cite{pr_144_708} and the applications of it began as early as 1980.\cite{prb_21_4656} What is new in the present work is that the kernel of Bethe-Salpeter equation is obtained consistently from the scGW calculation, whereas it is quite common even nowadays to use LDA approximation (often supplemented with scissor operator for the one-electron spectrum correction) to generate G and W, or, to replace LDA with G$_{0}$W$_{0}$/QSGW approximation which are either not self-consistent or (like LDA) not diagrammatic and, as a result, their connection with the Bethe-Salpeter equation is not clear. Another step forward is the elimination of the so called static approximation for W in the kernel of Bethe-Salpeter equation which still is in common use.

\section{Conclusions}
\label{concl}

In conclusion, two vertex corrected schemes to solve the Hedin's equations self-consistently have been applied to calculate the band gaps for a number of semiconductors/insulators. Undoubtedly, they both improve the results from scGW and QSGW calculations considerably. For this class of materials, the approach based on the physical polarizability (scheme D) results in slightly smaller band gaps than the conserving scheme B. However, it is hard to say decisively which scheme is better because the difference in their results is small and, in fact, is often less than the size of the electron-phonon effects, which are known only approximately.

Comparison with the experimental data suggests that one has to take the electron-phonon effects into consideration if one wants to enhance the predictive power of the theoretical approach.

This work was   supported by the U.S. Department of energy, Office of Science, Basic
Energy Sciences as a part of the Computational Materials Science Program. I thank Bartomeu Monserrat for pointing out the importance of the electron-phonon effects when comparing the theoretical and the experimental band gaps. Numerous discussions with Gabriel Kotliar are highly appreciated.


\begin{thebibliography}{33}
\expandafter\ifx\csname natexlab\endcsname\relax\def\natexlab#1{#1}\fi
\expandafter\ifx\csname bibnamefont\endcsname\relax
  \def\bibnamefont#1{#1}\fi
\expandafter\ifx\csname bibfnamefont\endcsname\relax
  \def\bibfnamefont#1{#1}\fi
\expandafter\ifx\csname citenamefont\endcsname\relax
  \def\citenamefont#1{#1}\fi
\expandafter\ifx\csname url\endcsname\relax
  \def\url#1{\texttt{#1}}\fi
\expandafter\ifx\csname urlprefix\endcsname\relax\def\urlprefix{URL }\fi
\providecommand{\bibinfo}[2]{#2}
\providecommand{\eprint}[2][]{\url{#2}}

\bibitem[{\citenamefont{{W.~Kohn and L.~J.~Sham}}(1965)}]{pr_140_1133}
\bibinfo{author}{\bibnamefont{{W.~Kohn and L.~J.~Sham}}},
  \bibinfo{journal}{Phys.~Rev.} \textbf{\bibinfo{volume}{140}},
  \bibinfo{pages}{A1133} (\bibinfo{year}{1965}).

\bibitem[{\citenamefont{{L.~Hedin}}(1965)}]{pr_139_A796}
\bibinfo{author}{\bibnamefont{{L.~Hedin}}}, \bibinfo{journal}{Phys.~Rev.}
  \textbf{\bibinfo{volume}{139}}, \bibinfo{pages}{A796} (\bibinfo{year}{1965}).

\bibitem[{\citenamefont{{G.~Strinati, H.~J.~Mattausch, and
  W.~Hanke}}(1982)}]{prb_25_2867}
\bibinfo{author}{\bibnamefont{{G.~Strinati, H.~J.~Mattausch, and W.~Hanke}}},
  \bibinfo{journal}{Phys.~Rev.~B} \textbf{\bibinfo{volume}{25}},
  \bibinfo{pages}{2867} (\bibinfo{year}{1982}).

\bibitem[{\citenamefont{{M.~S.~Hybertsen and S.~G.~Louie}}(1986)}]{prb_34_5390}
\bibinfo{author}{\bibnamefont{{M.~S.~Hybertsen and S.~G.~Louie}}},
  \bibinfo{journal}{Phys.~Rev.B} \textbf{\bibinfo{volume}{34}},
  \bibinfo{pages}{5390} (\bibinfo{year}{1986}).

\bibitem[{\citenamefont{{R.~W.~Godby, M.~Schl\"{u}ter, and
  L.~J.~Sham}}(1988)}]{prb_37_10159}
\bibinfo{author}{\bibnamefont{{R.~W.~Godby, M.~Schl\"{u}ter, and L.~J.~Sham}}},
  \bibinfo{journal}{Phys.~Rev.~B} \textbf{\bibinfo{volume}{37}},
  \bibinfo{pages}{10159} (\bibinfo{year}{1988}).

\bibitem[{\citenamefont{{J.~E.~Northrup, M.~S.~Hybertsen, and
  S.~G.~Louie}}(1987)}]{prl_59_819}
\bibinfo{author}{\bibnamefont{{J.~E.~Northrup, M.~S.~Hybertsen, and
  S.~G.~Louie}}}, \bibinfo{journal}{Phys.~Rev.~Lett.}
  \textbf{\bibinfo{volume}{59}}, \bibinfo{pages}{819} (\bibinfo{year}{1987}).

\bibitem[{\citenamefont{{M.~Shishkin, M.~Marsman, and
  G.~Kresse}}(2007)}]{prl_99_246403}
\bibinfo{author}{\bibnamefont{{M.~Shishkin, M.~Marsman, and G.~Kresse}}},
  \bibinfo{journal}{Phys.~Rev.~Lett.} \textbf{\bibinfo{volume}{99}},
  \bibinfo{pages}{246403} (\bibinfo{year}{2007}).

\bibitem[{\citenamefont{{W.~Ku and A.~G.~Eguiluz}}(2002)}]{prl_89_126401}
\bibinfo{author}{\bibnamefont{{W.~Ku and A.~G.~Eguiluz}}},
  \bibinfo{journal}{Phys.~Rev.~Lett.} \textbf{\bibinfo{volume}{89}},
  \bibinfo{pages}{126401} (\bibinfo{year}{2002}).

\bibitem[{\citenamefont{{T.~Kotani and M.~van~Schilfgaarde,
  S.~V.~Faleev}}(2007)}]{prb_76_165106}
\bibinfo{author}{\bibnamefont{{T.~Kotani and M.~van~Schilfgaarde,
  S.~V.~Faleev}}}, \bibinfo{journal}{Phys.~Rev.B}
  \textbf{\bibinfo{volume}{76}}, \bibinfo{pages}{165106}
  (\bibinfo{year}{2007}).

\bibitem[{\citenamefont{{A.~Kutepov, K.~Haule, S.~Y.~Savrasov, and
  G.~Kotliar}}(2012)}]{prb_85_155129}
\bibinfo{author}{\bibnamefont{{A.~Kutepov, K.~Haule, S.~Y.~Savrasov, and
  G.~Kotliar}}}, \bibinfo{journal}{Phys.~Rev.~B} \textbf{\bibinfo{volume}{85}},
  \bibinfo{pages}{155129} (\bibinfo{year}{2012}).

\bibitem[{\citenamefont{{R.~T.~M.~Ummels, P.~A.~Bobbert, and
  W.~van~Haeringen}}(1998)}]{prb_57_11962}
\bibinfo{author}{\bibnamefont{{R.~T.~M.~Ummels, P.~A.~Bobbert, and
  W.~van~Haeringen}}}, \bibinfo{journal}{Phys.~Rev.~B}
  \textbf{\bibinfo{volume}{57}}, \bibinfo{pages}{11962} (\bibinfo{year}{1998}).

\bibitem[{\citenamefont{{L.~Reining, V.~Olevano, A.~Rubio, and
  G.~Onida}}(2002)}]{prl_88_066404}
\bibinfo{author}{\bibnamefont{{L.~Reining, V.~Olevano, A.~Rubio, and
  G.~Onida}}}, \bibinfo{journal}{Phys.~Rev.~Lett.}
  \textbf{\bibinfo{volume}{88}}, \bibinfo{pages}{066404}
  (\bibinfo{year}{2002}).

\bibitem[{\citenamefont{{A.~Gr\"{u}neis, G.~Kresse, Y.~Hinuma, and
  F.~Oba}}(2014)}]{prl_112_096401}
\bibinfo{author}{\bibnamefont{{A.~Gr\"{u}neis, G.~Kresse, Y.~Hinuma, and
  F.~Oba}}}, \bibinfo{journal}{Phys.~Rev.~Lett.}
  \textbf{\bibinfo{volume}{112}}, \bibinfo{pages}{096401}
  (\bibinfo{year}{2014}).

\bibitem[{\citenamefont{{A.~L.~Kutepov}}(2016)}]{prb_94_155101}
\bibinfo{author}{\bibnamefont{{A.~L.~Kutepov}}},
  \bibinfo{journal}{Phys.~Rev.~B} \textbf{\bibinfo{volume}{94}},
  \bibinfo{pages}{155101} (\bibinfo{year}{2016}).

\bibitem[{\citenamefont{{R.~Del~Sole, L.~Reining,
  R.~W.~Godby}}(1994)}]{prb_49_8024}
\bibinfo{author}{\bibnamefont{{R.~Del~Sole, L.~Reining, R.~W.~Godby}}},
  \bibinfo{journal}{Phys.~Rev.~B} \textbf{\bibinfo{volume}{49}},
  \bibinfo{pages}{8024} (\bibinfo{year}{1994}).

\bibitem[{\citenamefont{{F.~Bruneval, F.~Sottile, V.~Olevano, R.~Del~Sole, and
  L.~Reining}}(2005)}]{prl_94_186402}
\bibinfo{author}{\bibnamefont{{F.~Bruneval, F.~Sottile, V.~Olevano,
  R.~Del~Sole, and L.~Reining}}}, \bibinfo{journal}{Phys.~Rev.~Lett.}
  \textbf{\bibinfo{volume}{94}}, \bibinfo{pages}{186402}
  (\bibinfo{year}{2005}).

\bibitem[{\citenamefont{{F.~J.~Himpsel, J.~A.~Knapp, J.~A.~VanVechten, and
  D.~E.~Eastman}}(1979)}]{prb_20_624}
\bibinfo{author}{\bibnamefont{{F.~J.~Himpsel, J.~A.~Knapp, J.~A.~VanVechten,
  and D.~E.~Eastman}}}, \bibinfo{journal}{Phys.~Rev.~B}
  \textbf{\bibinfo{volume}{20}}, \bibinfo{pages}{624} (\bibinfo{year}{1979}).

\bibitem[{\citenamefont{{P.~Lautenschlager, M.~Garriga, S.~Logothetidis, and
  M.~Cardona}}(1987)}]{prb_35_9174}
\bibinfo{author}{\bibnamefont{{P.~Lautenschlager, M.~Garriga, S.~Logothetidis,
  and M.~Cardona}}}, \bibinfo{journal}{Phys.~Rev.~B}
  \textbf{\bibinfo{volume}{35}}, \bibinfo{pages}{9174} (\bibinfo{year}{1987}).

\bibitem[{\citenamefont{{A.~Mang, K.~Reimann, St.~R\"{u}benacke, and
  M.~Steube}}(1996)}]{prb_53_16283}
\bibinfo{author}{\bibnamefont{{A.~Mang, K.~Reimann, St.~R\"{u}benacke, and
  M.~Steube}}}, \bibinfo{journal}{Phys.~Rev.~B} \textbf{\bibinfo{volume}{53}},
  \bibinfo{pages}{16283} (\bibinfo{year}{1996}).

\bibitem[{\citenamefont{{B.~Monserrat and R.~J.~Needs}}(2014)}]{prb_89_214304}
\bibinfo{author}{\bibnamefont{{B.~Monserrat and R.~J.~Needs}}},
  \bibinfo{journal}{Phys.~Rev.~B} \textbf{\bibinfo{volume}{89}},
  \bibinfo{pages}{214304} (\bibinfo{year}{2014}).

\bibitem[{\citenamefont{{G.~Antonius, S.~Ponce, P.~Boulanger, M.~Cote, and
  X.~Gonze}}(2014)}]{prl_112_215501}
\bibinfo{author}{\bibnamefont{{G.~Antonius, S.~Ponce, P.~Boulanger, M.~Cote,
  and X.~Gonze}}}, \bibinfo{journal}{Phys.~Rev.~Lett.}
  \textbf{\bibinfo{volume}{112}}, \bibinfo{pages}{215501}
  (\bibinfo{year}{2014}).

\bibitem[{\citenamefont{{M.~Cardona, M.~L.~W.~Thewalt}}(2005)}]{rmp_77_1173}
\bibinfo{author}{\bibnamefont{{M.~Cardona, M.~L.~W.~Thewalt}}},
  \bibinfo{journal}{Rev.~Mod.~Phys.} \textbf{\bibinfo{volume}{77}},
  \bibinfo{pages}{1173} (\bibinfo{year}{2005}).

\bibitem[{\citenamefont{{B.~Monserrat}}(2016)}]{prb_93_100301}
\bibinfo{author}{\bibnamefont{{B.~Monserrat}}}, \bibinfo{journal}{Phys.~Rev.~B}
  \textbf{\bibinfo{volume}{93}}, \bibinfo{pages}{100301}
  (\bibinfo{year}{2016}).

\bibitem[{\citenamefont{{G.~Strinati}}(1988)}]{rdnc_11_1}
\bibinfo{author}{\bibnamefont{{G.~Strinati}}},
  \bibinfo{journal}{Riv.~del~Nuovo~Cimento} \textbf{\bibinfo{volume}{11}},
  \bibinfo{pages}{1} (\bibinfo{year}{1988}).

\bibitem[{\citenamefont{{W.~Chen and A.~Pasquarello}}(2015)}]{prb_92_041115}
\bibinfo{author}{\bibnamefont{{W.~Chen and A.~Pasquarello}}},
  \bibinfo{journal}{Phys.~Rev.~B} \textbf{\bibinfo{volume}{92}},
  \bibinfo{pages}{041115} (\bibinfo{year}{2015}).

\bibitem[{\citenamefont{{C.~Friedrich, A.~Schindlmayr, and S.~Bl\"{u}gel,
  T.~Kotani}}(2006)}]{prb_74_045104}
\bibinfo{author}{\bibnamefont{{C.~Friedrich, A.~Schindlmayr, and S.~Bl\"{u}gel,
  T.~Kotani}}}, \bibinfo{journal}{Phys.~Rev.~B} \textbf{\bibinfo{volume}{74}},
  \bibinfo{pages}{045104} (\bibinfo{year}{2006}).

\bibitem[{\citenamefont{{C.~Friedrich, M.~C.~M\"{u}ller, and
  S.~Bl\"{u}gel}}(2011{\natexlab{a}})}]{prb_83_081101}
\bibinfo{author}{\bibnamefont{{C.~Friedrich, M.~C.~M\"{u}ller, and
  S.~Bl\"{u}gel}}}, \bibinfo{journal}{Phys.~Rev.~B}
  \textbf{\bibinfo{volume}{83}}, \bibinfo{pages}{081101}
  (\bibinfo{year}{2011}{\natexlab{a}}).

\bibitem[{\citenamefont{{C.~Friedrich, M.~C.~M\"{u}ller, and
  S.~Bl\"{u}gel}}(2011{\natexlab{b}})}]{prb_84_039906}
\bibinfo{author}{\bibnamefont{{C.~Friedrich, M.~C.~M\"{u}ller, and
  S.~Bl\"{u}gel}}}, \bibinfo{journal}{Phys.~Rev.~B}
  \textbf{\bibinfo{volume}{84}}, \bibinfo{pages}{039906(E)}
  (\bibinfo{year}{2011}{\natexlab{b}}).

\bibitem[{\citenamefont{{H.~Jiang, and P.~Blaha}}(2016)}]{prb_93_115203}
\bibinfo{author}{\bibnamefont{{H.~Jiang, and P.~Blaha}}},
  \bibinfo{journal}{Phys.~Rev.~B} \textbf{\bibinfo{volume}{93}},
  \bibinfo{pages}{115203} (\bibinfo{year}{2016}).

\bibitem[{\citenamefont{{D.~Nabok, A.~Gulans, and
  C.~Draxl}}(2016)}]{prb_94_035118}
\bibinfo{author}{\bibnamefont{{D.~Nabok, A.~Gulans, and C.~Draxl}}},
  \bibinfo{journal}{Phys.~Rev.~B} \textbf{\bibinfo{volume}{94}},
  \bibinfo{pages}{035118} (\bibinfo{year}{2016}).

\bibitem[{\citenamefont{{W.~A.~Caliebe, J.~A.~Soininen, E.~L.~Shirley,
  C.~-C.~Kao, K.~H\"{a}m\"{a}l\"{a}inen}}(2000)}]{prl_84_3907}
\bibinfo{author}{\bibnamefont{{W.~A.~Caliebe, J.~A.~Soininen, E.~L.~Shirley,
  C.~-C.~Kao, K.~H\"{a}m\"{a}l\"{a}inen}}}, \bibinfo{journal}{Phys.~Rev.~Lett.}
  \textbf{\bibinfo{volume}{84}}, \bibinfo{pages}{3907} (\bibinfo{year}{2000}).

\bibitem[{\citenamefont{{L.~J.~Sham and T.~M.~Rice}}(1966)}]{pr_144_708}
\bibinfo{author}{\bibnamefont{{L.~J.~Sham and T.~M.~Rice}}},
  \bibinfo{journal}{Phys.~Rev.} \textbf{\bibinfo{volume}{144}},
  \bibinfo{pages}{708} (\bibinfo{year}{1966}).

\bibitem[{\citenamefont{{W.~Hanke and L.~J.~Sham}}(1980)}]{prb_21_4656}
\bibinfo{author}{\bibnamefont{{W.~Hanke and L.~J.~Sham}}},
  \bibinfo{journal}{Phys.~Rev.~B} \textbf{\bibinfo{volume}{21}},
  \bibinfo{pages}{4656} (\bibinfo{year}{1980}).

\end{thebibliography}

\end{document}